\documentclass[12pt]{iopart}
\usepackage{amsmath}
\usepackage{graphicx}
\usepackage{xcolor}
\usepackage[pdftex, pdftitle={Article}, pdfauthor={Author}]{hyperref}
\usepackage{multirow}
\usepackage{subcaption}

\newcommand{\tsub}{\textsubscript}
\newcommand{\tsup}{\textsuperscript}

\begin{document}

\title[Phase diagram of strongly-coupled Rashba systems]{Phase diagram of strongly-coupled Rashba systems}

%
%
%
%

\author{B. K. Nally\tsup{1} and P. M. R. Brydon\tsup{1*}}
\address{{}\tsup{1}Department of Physics and MacDiarmid Institute for Advanced Materials and Nanotechnology, Univeristy of Otago, P.O. Box 56, Dunedin 9054, New Zealand}
\ead{\mailto{philip.brydon@otago.ac.nz}}

\vspace{10pt}
\begin{indented}
\item[]{14 May 2024}
\end{indented}

\begin{abstract}
Motivated by the recent discovery of a possible field-mediated parity switch within the superconducting state of CeRh\tsub{2}As\tsub{2} [Khim et al., Science \textbf{373}, 1012 (2021)], we thoroughly investigate the dependence of the superconducting state of a strongly-coupled Rashba mono- and bilayer on internal parameters and an applied magnetic field. The role of interlayer pairing, spin orbit coupling, doping rate and applied magnetic field and their interplay was examined numerically at low temperature within a \emph{t-J-}like model, uncovering complex phase diagrams and transitions between superconducting states with different symmetry.
\end{abstract}

\section{Introduction}\label{sec:intro}

Inversion symmetry is an important symmetry in the phenomenology of superconductivity, as its presence requires that the electrons pair in either a singlet or triplet state. In non-centrosymmetric (NCS) materials, however, the breaking of inversion symmetry means that this distinction between singlet and triplet no longer exists. NCS superconductors are distinguished by the lifting of spin degeneracy by the SOC, and the existence of a pairing state with both singlet and triplet components \cite{gorkov:2001, frigeri:2004, samokhin:2004}.  Of more recent interest are locally noncentrosymmetric (LNCS) superconductors (SCs), where global inversion symmetry is retained but with a sublattice structure where the atomic sites are not centres of inversion. Despite being inversion symmetric, this sublattice structure is nevertheless predicted to lead to behavior similar to NCS systems, such as enhanced Pauli limiting fields due to SOC and singlet/triplet mixing \cite{yoshida:2012,sigrist:2014,nakamura:2017,maruyama:2012,fischer:2011,maruyama:2012,fischer:2023}. LNCS materials include artificial superlattices and some heavy fermion, cuprate and pnictide SCs \cite{youn:2012,fischer:2023,kibune:2022}.

Whilst LNCS systems do not necessarily always realize unconventional pairing, the conventional states they host also display unusual character, such as enhanced upper critical fields \cite{khim:2021,fischer:2023}. They also may host an unconventional state unique to LNCS systems -- a ``staggered'' state where the SC wavefunction changes sign between the sublattice layers, which is almost impervious to magnetic fields applied perpendicular to the layers \cite{sigrist:2014,khim:2021,mockli:2021}. Recently discovered CeRh\tsub{2}As\tsub{2} is an LNCS SC which is suspected to undergo a field-mediated transition between this staggered state and the conventional uniform state \cite{khim:2021,schertenleib:2021}. The relative strengths of the SOC, which couples singlet and triplet pairings, and intersublattice-hopping, which mutes the effect of the local noncentrosymmetry, have been shown to be a strong indicator of staggered superconducting states in models of LNCS systems \cite{sigrist:2014,maruyama:2012}. Whilst these materials are generally well-studied, this has been mostly focused on the weak coupling limit behavior of these systems \cite{medhi:2009,nogaki:2020}.

Herein we present a thorough investigation into a Rashba monolayer and bilayer in the strong coupling limit using a \emph{t-J-}like model, utilizing a slave boson approach.  The effect of internal and external parameters - doping, SOC strength, interlayer hopping amplitude, and magnetic field strength - on the nature of the SC state are investigated at low temperature. We first investigate a monolayer in order to study the effects of the SOC, and find that the SOC tends to drive a transition from a $d_{x^2-y^2}$-wave dominant mixed state to an extended-$s$-wave dominant pairing. Then, we stack two monolayers and couple them with an interlayer hopping (ILH). In this bilayer, the SOC drives a transition from $d$-wave to $s$-wave pairing as in the monolayer. We find that the ILH also favours the $s$-wave state. The odd-parity states, corresponding to staggered $s$-wave or $d$-wave, are found to exist even at zero field strength, and also generically appear as the field is increased.  We then examine the energetics of the system to investigate the stabilizing factors for the competing states. 
\section{Monolayer system}\label{sec:mono}

\subsection{Model}

We first consider the noncentrosymmetric monolayer, in point group $C_{4v}$, which is described by the Hamiltonian
\begin{align}
    H =&  H_\text{int} + H_0,\label{eq:H}\\
    H_0 =& \sum_{\langle i,j \rangle}\sum_{s,s'}  -tc^\dagger_{i,s}(\sigma^0)_{ss'}c_{j,s'} + i\lambda c^\dagger_{i,s}\hat{r}_{ij}\cdot\left(\sigma^x\hat{y}-\sigma^y\hat{x}\right)_{s,s'}c_{j,s'}\label{eq:H0}\\
    H_\text{int} =& U\sum_{i} n_{i,\uparrow}n_{i,\downarrow}.
    \end{align} 
The normal-state Hamiltonian $H_0$ is characterized by a nearest-neighbour hopping $t$ on a square lattice and a Rashba SOC of strength $\lambda$, with the latter allowed by the broken inversion symmetry. In~\eref{eq:H0} the $\sigma^\mu$ matrices denote the Pauli spin matrices, $\hat{r}_{ij}$ is the unit vector between sites $i$ and $j$, and $\langle i,j \rangle$ denotes summation over nearest neighbours only. The interaction term is a Hubbard on-site repulsion of strength $U$. Here $c_{j,s}$ and $c^\dagger_{j,s}$ have the usual meaning of annihilation and creation operators for spin-$s$ electrons on site $j$, and $n_{j,s}$ is the corresponding number operator.

We analyze the Hamiltonian~\eref{eq:H} in the strong coupling limit, where $U$ far exceeds the bandwidth of the normal-state Hamiltonian. We adopt the approximation that this excludes doubly-occupied sites except as virtual processes. Projected onto the subspace excluding double occupancy, we have the effective Hamiltonian
\begin{align}
    H_\text{eff} &= \tilde{H_0} + H_J\label{eq:effH}
\end{align}
where $\tilde{H_0}$ is as in~\eref{eq:H0} with the electron operators replaced by their projections in the no-double-occupation space and $H_J$ is obtained via canonical transformation~\cite{fazekas}: 
\begin{align}
\begin{split}
        H_J = \sum_{\langle i,j \rangle} &J\left[ \mathbf{S}_i\cdot \mathbf{S}_j - \frac{n_in_j}{4} \right] + D (\hat{\mathbf{z}}\times\mathbf{\hat{e}_{ij}})\cdot  (\mathbf{S}_i \times \mathbf{S}_j) \\
        + &J^\prime \left[\alpha_{|j-i|}\left( S_i^xS_j^x -S_i^yS_j^y\right) - S_i^zS_j^z  - \frac{n_in_j}{4} \right]\label{eq:HJ}
\end{split}\end{align}
where $J=4t^2/U$ is the Heisenberg coupling, $J^\prime=-4\lambda^2/U$ the compass interaction and $D=8t\lambda/U$ the Dzyaloshinskii-Moriya (DM) coupling. Note that the compass interaction includes the direction-dependent factor $\alpha_{|j-i|}$ where $\alpha_{x}=-1$ and $\alpha_y=1$. The DM interaction depends non-trivially on the hopping direction, $\hat{\mathbf{e}}_{ij}$.

We use the auxiliary boson method \cite{barnes:1976,coleman:1984} to develop a mean-field (MF) theory for the effective Hamiltonian. The electron operators are replaced as $c^\dagger_{i,\sigma} \rightarrow f^\dagger_{i,\sigma} b_i$, where the $f_\sigma$ and $b$ are annihilation operators for the fermionic spinons and bosonic holons, respectively. At the MF level, the no-double-occupancy condition is enforced on the average with
\begin{align}
	1 = \sum_{i,\sigma} \langle f^\dagger_{i\sigma}f_{i\sigma} \rangle + \langle b^\dagger_ib_i\rangle = \sum_{i,\sigma} \langle f^\dagger_{i\sigma}f_{i\sigma} \rangle + \delta.\label{q:nodoublee}
\end{align}
where the density of holons is equal to the doping, i.e. $\langle b^\dagger_ib_i\rangle = \delta$ \cite{ogata:2008}. Following the usual approach, we assume that the holons are condensed at low temperatures \cite{inaba:1996, yamase:2011}, and so we can use the MF approximation $\langle b_i \rangle = \sqrt{\delta}$. This auxiliary boson technique arises as a natural MF treatment of strongly coupled systems, which are well described by a resonating valence bond (RVB) state background \cite{anderson:1987rvb,anderson:1987,baskaran:1987}.

We proceed to apply the MF approximation to the projected Hamiltonian as follows: we replace the electron annihilation and creation operators by the spinon and holon operators, with the holon operators further replaced by their expectation value, e.g. $ b_i^\dagger b_j \approx \langle b_i^\dagger \rangle \langle b_j \rangle = \delta$. We further decouple the nearest-neighbour interactions in the particle-particle and particle-hole channels, introducing the MF amplitudes
\begin{align}
	\chi^\mu_a &= \langle f_{is_1}^\dagger (\sigma^\mu)_{s_1s_2} f_{i+as_2} \rangle \\
	\Delta^{\mu}_a &= \langle f_{i+as_1} (i\sigma^\mu\sigma^y)_{s_1s_2} f_{is_2} \rangle
\end{align}
The normal-state amplitudes $\chi$ and $\chi^\text{SOC}$ are defined as
\begin{align}
\chi &= \chi_{\pm x}^0 = \chi_{\pm y}^0 \label{eq:chi} \\
\chi^\text{SOC} &= \pm \chi^y_{\pm x} = \mp \chi^x_{\pm y}. \label{eq:chisoc}     
\end{align}

The renormalization or bond parameters $\chi$ and $\chi^\text{SOC}$ are a measure of the probability of spin-preserving and spin-flipping hopping between neighbouring sites, respectively. This can alternately be thought of as singlet bonds and opposite-spin triplet bonds; e.g., an electron undergoing a spin-preserving hopping requires an opposite spin partner in the neighbouring site, effectively creating a singlet bond. The amplitudes $\chi$ and $\chi^\text{SOC}$ renormalize the nearest-neighbour hopping and the Rashba SOC in the noninteracting Hamiltonian, respectively. Doping away from the half-filled RVB background state and adding SOC allows spin mobility and spin-flip processes, and hence singlet/triplet mixed-state superconductivity and multiple bond parameters.

For now we restrict our attention for the gap amplitudes $\Delta^\mu_a$ to the $A_1$ and $B_1$ irreducible representations (irreps) of the point group $C_{4v}$, as these contain singlet pairing states which are favoured by the nearest-neighbour interaction in \eref{eq:HJ}; the pairing amplitudes for other irreps are small or vanishing. For the $A_1$ irrep, the gap amplitudes are
\begin{align}
\Delta^s &= \Delta^0_{\pm x} = \Delta^0_{\pm y}\\
\Delta^{t} &= \pm \Delta^y_{\pm x} = \mp \Delta^x_{\pm y}
\end{align}
while for the $B_1$ irrep we have
\begin{align}
\Delta^s &= \Delta^0_{\pm x} = -\Delta^0_{\pm y}\\
\Delta^{t} &= \pm \Delta^y_{\pm x} = \pm \Delta^x_{\pm y}.
\end{align}
The pairing potentials in momentum space are listed in \tref{tab:monoOPs}. Note that the  singlet $A_1$ component corresponds to an extended $s$-wave pairing, whereas the singlet $B_1$ is $d_{x^2-y^2}$-wave. The MF amplitudes are determined by solving the self-consistency equations, and the stable superconducting state is determined from the free energy. In our numerical calculations we utilize a lattice of $200 \times 200$ ${\bf k}$-points,  and the temperature is taken to be $T=0.001J$.

\begin{table}
    \centering
    \begin{tabular}{|c|c|c|}
    \hline
        $C_{4v}$       & singlet      &  triplet   \\\hline
        $A_{1}$        & $[\cos(k_xa)+\cos(k_ya)]\sigma^0$      & $\sin(k_ya)\sigma^x-\sin(k_xa)\sigma^y$ \\
        $B_{1}$        &$[\cos(k_xa)-\cos(k_ya)]\sigma^0$       & $\sin(k_xa)\sigma^y+\sin(k_ya)\sigma^x$  \\\hline
    \end{tabular}
    \caption{Irrep assignments for the momentum and spin components of the monolayer MF parameters studied. Renormalization MF parameters belong to the totally symmetric $A_{1}$ irrep, and are proportional to $(f^\dagger_{\mathbf{k}s}(...)_{s,s'}f_{\mathbf{k}s'})$ where $(...)$ are table entries above. Superconducting gap functions are proportional to $(f_{\mathbf{-k}s}(...i\sigma^y)_{s,s'}f_{\mathbf{k}s'})$. Only functions possible with only nearest-neighbour interactions are included.}
    \label{tab:monoOPs}
\end{table}

\subsection{Results}

A monolayer model similar to the one considered here was studied in~\cite{farrell:2014} using the Gutzwiller approximation. These authors considered relatively weak spin-orbit coupling, and our results in this limit are in agreement with theirs.

\begin{figure}
    \centering
    \includegraphics[width=0.7\linewidth]{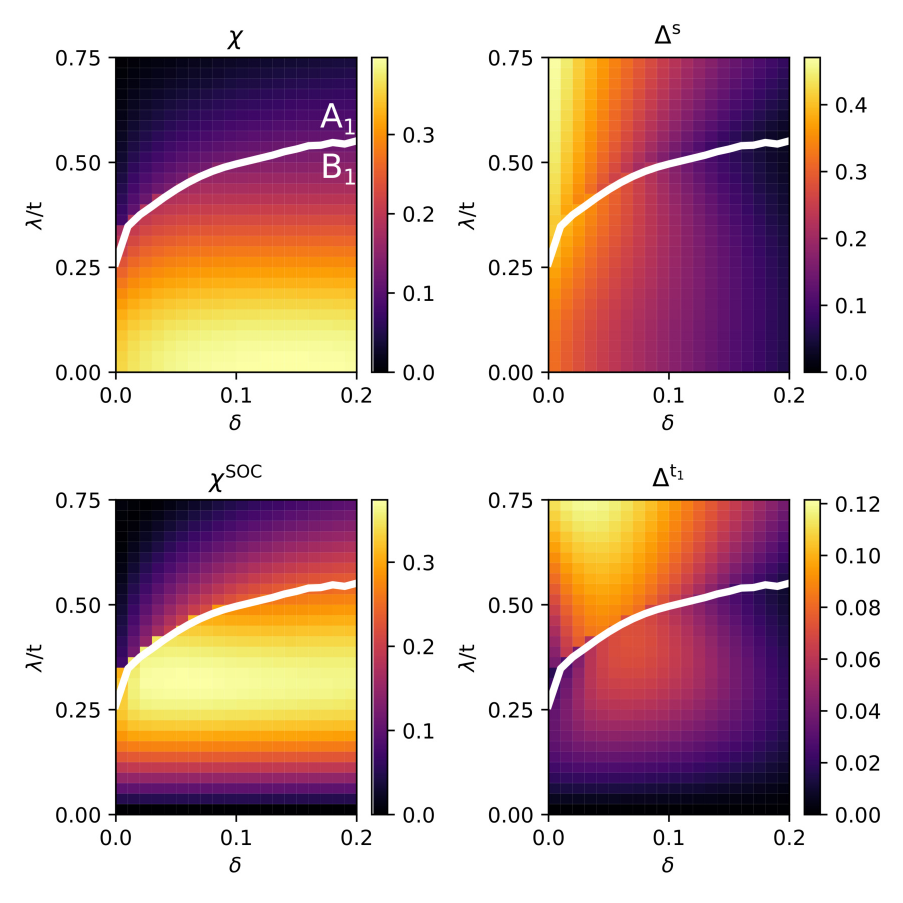}
    \caption{{ MF amplitudes for the monolayer system in as a function of hole doping $\delta$ and SOC strength $\lambda$ space. The white line indicates the boundary between the $A_1$ and $B_1$ superconducting phases. Note that the color ranges vary between plots for clarity.}}\label{fig:OP_mono_ld}
\end{figure}

In \fref{fig:OP_mono_ld} we show the variation of the MF parameters as a function of hole doping and SOC strength. Our main result here is that there is a transition between the $B_1$ and $A_1$ superconducting states with increasing SOC strength, and the transition line moves to higher SOC strengths with increasing doping. In both superconducting states the singlet pairing amplitude dominates over the triplet.

\begin{figure}[h]
    \centering
    \includegraphics[width=0.7\linewidth]{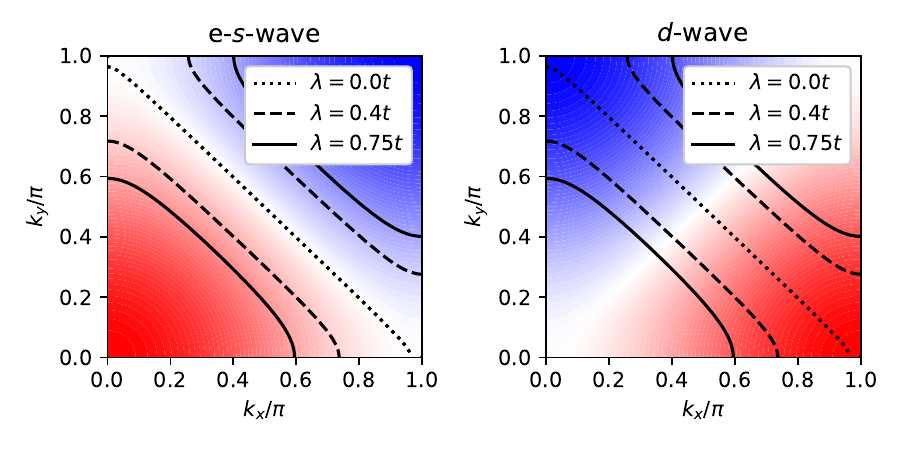}
    \caption{Fermi surface for various SOC strengths at hole doping $\delta=0.25$ overlaid on the different pairing functions. Nodes of the pairing functions lie on the white contours.}\label{fig:fs_all}
\end{figure}

The transition from the $d$-wave-dominated $B_1$ to extended-$s$-wave-dominated $A_1$ pairing state can be understood in terms of the lifting of the spin-degeneracy by the SOC, which results in two Fermi surfaces (FS) in the normal state (NS) as illustrated in \fref{fig:fs_all}. In the absence of SOC and weak doping, the FS passes close to the gap maxima of the $B_1$ state; in contrast, the FS lies close to the nodal line of the $A_1$ extended-$s$-wave state. Thus, the average gap opened by the $B_1$ state over the FS is much larger than that opened by the $A_1$, which energetically favours the former. Upon switching on the SOC, however, the spin-split FSs will tend to move away from the nodal line of the $A_1$ state, thus enhancing its stability relative to the $B_1$ state, and eventually ensuring that it has the lower free energy. As we increase the doping, the Fermi surface moves off the nodal line. Although the extended $s$-wave state can now open a full gap, it remains less stable than the $d$-wave. Moreover, it requires a larger SOC to stabilize, since the gap on one of the spin-split Fermi surfaces initially decreases as the SOC is switched on; only after it crosses the nodal line can the SOC stabilize the extended $s$-wave.

As for the subdominant triplet gaps, \fref{fig:OP_mono_ld} shows that the triplet amplitude is immediately enhanced upon the transition into the $A_1$ state. This likely reflects the fact that the $A_1$ triplet state is insensitive to the SOC, whereas the SOC is pair-breaking for the $B_1$ state, i.e. the $A_1$ triplet state only pairs quasiparticles in the same spin-split (``helicity'') bands. Although a $B_1$ triplet state with same-helicity pairing is possible, this requires $f$-wave pair amplitude and thus interactions beyond nearest neighbour \cite{sigrist:2,yokoyama:2007}.

As discussed in~\cite{farrell:2014}, the monolayer NCS hs nontrivial topological properties. We have not considered the topological character of the phases in our system, but this is potentially a fruitful direction for further work.
\section{Bilayer system}\label{sec:bi}

\subsection{Model}
The bilayer system can be considered as two copies of the monolayer with opposite sign of the SOC, which are coupled by an interlayer hopping (ILH) term $t_\perp$ \cite{maruyama:2012,yoshida:2012,schertenleib:2021,khim:2021}. The layer degree of freedom restores the inversion symmetry and takes the system to the point group $D_{4h}$, and we encode this degree of freedom in the $\eta^\mu$ Pauli matrices. The non-interacting part of the Hamiltonian is thus
\begin{align}
    H_0 =& \sum_{\langle i,j \rangle}\sum_{s,s',l,l'}  -tc^\dagger_{i,s,l}\left(\eta^0_{ll'}\sigma^0_{ss'} \right)c_{j,s',l'}+i\lambda c^\dagger_{i,s,l}\eta^z_{ll'}\hat{r}_{ij}\cdot\left(\sigma^x_{ss'}\hat{y}-\sigma^y_{ss'}\hat{x}\right)c_{j,s',l'}\notag\\
    &  + t_\perp \sum_j\sum_{s,s',l,l'}  c^\dagger_{j,s,l}  \left(\eta^x_{ll'}\sigma^0_{ss'}\right)c_{j,s',l'} \label{eq:H0_bi}
\end{align}
where $c_{j,s,l}$ is the annihilation operator for a spin-$s$ electron on layer $l=1,2$ of unit cell $j$. As in the monolayer case we include an on-site Hubbard interaction, which we assume to be the dominant energy scale.  Accordingly, we develop an effective theory excluding double-occupancy at each site of the bilayer. This gives effective nearest-neighbour interactions: in each layer this has the same form as \eref{eq:HJ}, but with opposite sign of the DM interaction. The ILH term gives rise to an additional Heisenberg interaction with exchange constant $J_\perp=4t_\perp^2/U$ between the spins in each layer of the same unit cell. 

The additional layer degree of freedom also increases the number of MFs, with both intralayer 
\begin{align}
	\chi^\mu_{a,l} &= \langle f_{i,s_1,l}^\dagger (\sigma^\mu)_{s_1s_2} f_{i+a,s_2,l} \rangle \\
	\Delta^{\mu}_{a,l} &= \langle f_{i+a,s_1,l} (i\sigma^\mu\sigma^y)_{s_1s_2} f_{i,s_2,l} \rangle
	    \end{align}
and interlayer amplitudes
\begin{align}
	\chi_\perp & = \langle f^\dagger_{isl}(\eta^x \otimes \sigma^0)_{ss'll'}f_{is'l'} \rangle\\
	\Delta_\perp & = \langle f_{isl}(\eta^x \otimes \sigma^0 i\sigma^y)_{ss'll'}f_{is'l'} \rangle
\end{align} 
The normal-state intralayer amplitudes have the in-plane variation as defined in \eref{eq:chi} and \eref{eq:chisoc}, but $\chi^{SOC}$ changes sign across the layers. For the pairing amplitudes, we focus on the irreps of the $D_{4h}$ point group which have nearest-neighbour spin-singlet pairing, namely the $A_{1g}$, $A_{2u}$, $B_{1g}$, and $B_{2u}$ irreps. The two intralayer gap amplitudes in the $A_{1g}$ irrep are
\begin{align}
\Delta^s &= \Delta^0_{\pm x,l=1,2} = \Delta^0_{\pm y,l=1,2}\\
\Delta^{t} &= \pm \Delta^y_{\pm x,1} = \mp \Delta^x_{\pm y,1} = \mp \Delta^y_{\pm x,2} = \pm \Delta^x_{\pm y,2}
\end{align}
The intralayer amplitudes are essentially the same as those defining the $A_1$ irrep of the monolayer, and with $s$-wave-like dominant pairing and with the subdominant triplet amplitudes $\Delta^t < \Delta^s$ reversing sign between each layer. There is also an interlayer gap amplitude $\Delta_\perp$, which is vanishing in the other irreps we consider here. For the $A_{2u}$ irrep, the singlet gap amplitude reverses sign across the layers whilst the triplet gap amplitude does not. Explicitly, the intralayer amplitudes are defined  
\begin{align}
\Delta^s &= \Delta^0_{\pm x,1} = \Delta^0_{\pm y,1}= -\Delta^0_{\pm x,2} = -\Delta^0_{\pm y,2}\\
\Delta^{t} &= \pm \Delta^y_{\pm x,l=1,2} = \mp \Delta^x_{\pm y,l=1,2}.
\end{align}
Similarly, the $B_{1g}$ and $B_{2u}$ states have intralayer amplitudes which are the same as in the $B_1$ irrep of the monolayer, but where the $d$-wave-like dominant singlet and subdominant triplet components reverse sign between the layers, respectively. That is, for the $B_{1g}$ we have 
\begin{align}
\Delta^s &= \Delta^0_{\pm x,l=1,2} = -\Delta^0_{\pm y,l=1,2}\\
\Delta^{t} &= \pm \Delta^y_{\pm x,1} = \pm \Delta^x_{\pm y,1} = \mp \Delta^y_{\pm x,2} = \mp \Delta^x_{\pm y,2}
\end{align}
whereas for the $B_{2u}$ 
\begin{align}
\Delta^s &= \Delta^0_{\pm x,1} = -\Delta^0_{\pm y,1}= -\Delta^0_{\pm x,2} = \Delta^0_{\pm y,2}\\
\Delta^{t} &= \pm \Delta^y_{\pm x,l=1,2} = \pm \Delta^x_{\pm y,l=1,2}
\end{align}
\Tref{tab:biOPs} shows the momentum-space form of the gap amplitudes for each irrep of the bilayer system. Note that since the interlayer pairing interaction acts only between sites in the same unit cell, the interlayer SC parameter is restricted to $s$-wave.

\begin{table*}
    \centering
    \begin{tabular}{|cc||c|c|c|c|}
    \hline
        &&\multicolumn{2}{c|}{intralayer} &\multicolumn{2}{c|}{interlayer}\\\hline
        $C_{4h}$& $D_{4h}$& singlet & triplet  & singlet & triplet   \\\hline
        \multirow{2}{*}{$A_g$}&$A_{1g}$& $\eta^0[\cos(k_xa)+\cos(k_ya)]\sigma^0$               & $\eta^z[\sin(k_ya)\sigma^x-\sin(k_xa)\sigma^y]$ & $\eta^x\sigma^0$& \\
                              &$A_{2g}$&                                        & $\eta^z[\sin(k_xa)\sigma^x+\sin(k_ya)\sigma^y]$ &                 & \\
        \multirow{2}{*}{$A_u$}&$A_{1u}$&                                        & $\eta^0[\sin(k_xa)\sigma^x+\sin(k_ya)\sigma^y]$ &                 & $\eta^y\sigma^z$\\
                              &$A_{2u}$& $\eta^z[\cos(k_xa)+\cos(k_ya)]\sigma^0$               & $\eta^0[\sin(k_ya)\sigma^x-\sin(k_xa)\sigma^y]$ & & \\
        \multirow{2}{*}{$B_g$}&$B_{1g}$&$\eta^0[\cos(k_xa)-\cos(k_ya)]\sigma^0$ & $\eta^z[\sin(k_xa)\sigma^y+\sin(k_ya)\sigma^x]$ &                 & \\
                              &$B_{2g}$&                                        & $\eta^z[\sin(k_xa)\sigma^x-\sin(k_ya)\sigma^y]$ &                 & \\
        \multirow{2}{*}{$B_u$}&$B_{1u}$&                                        & $\eta^0[\sin(k_xa)\sigma^x-\sin(k_ya)\sigma^y]$ &                 & \\
                              &$B_{2u}$&$\eta^z[\cos(k_xa)-\cos(k_ya)]\sigma^0$ & $\eta^0[\sin(k_xa)\sigma^y+\sin(k_ya)\sigma^x]$ &                 & \\\hline
    \end{tabular}
    \caption{Conditions on each MF parameter for all irreducible representations for pairing in the bilayer. Renormalization parameters follow the same rules as the totally symmetric $A_{1g}/A_{g}$ gap.}
    \label{tab:biOPs}
\end{table*}

\subsection{$d$-wave to extended-$s$-wave transition}

\begin{figure}
    \centering
    \includegraphics[width=\linewidth]{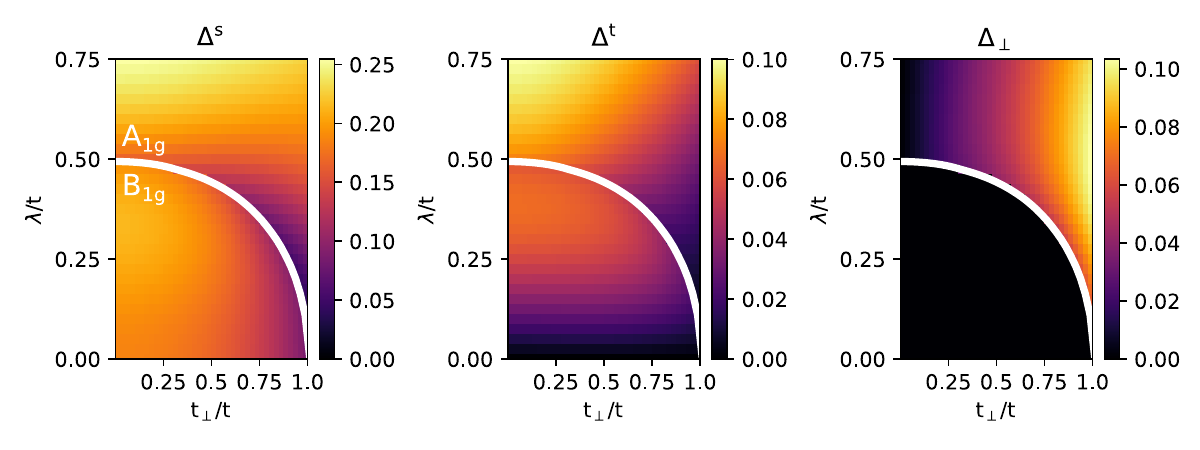}
    \caption{Interlayer MF parameter amplitudes for the bilayer system in $t_\perp-\lambda$ space for $\delta=0.1$, irrep of mixed state realized is indicated. Note that the color ranges vary between plots for clarity.}\label{fig:full_pd_tl}
\end{figure}

The phase diagram in $\lambda-t_\perp$ space can be seen in \fref{fig:full_pd_tl}. The odd states were not found to be favoured at any point in the chosen parameter ranges. The intralayer renormalization parameters were almost independent of $t_\perp$ and the ILH renormalization was found to be almost independent of $\lambda$, all with almost imperceptible change across the $B_{1g}$ to $A_{1g}$ transition. As such only the SC gaps are shown.

The stabilizing effect of the interlayer pairing for the $A_{1g}$ irrep is evident for non-zero $t_\perp$. Increasing the ILH has the effect of shifting the $A_{1g}-B_{1g}$ transition line to lower values of $\lambda$. This appears to be driven primarily by the interlayer singlet in the $A_{1g}$, which reaches a comparable amplitude to the dominant intralayer singlet pairing above about $t_\perp>0.8t$ and intermediate SOC strength. For ILH strengths $t_\perp> t$, the $B_{1g}$ is completely suppressed in favour of the $A_{1g}$, as the ILH also pushes the FSs outward -- similar to the effect the SOC had on the FSs which favoured the extended $s$-wave state in the monolayer.

\begin{figure}
\centering
    \includegraphics[width=0.8\linewidth]{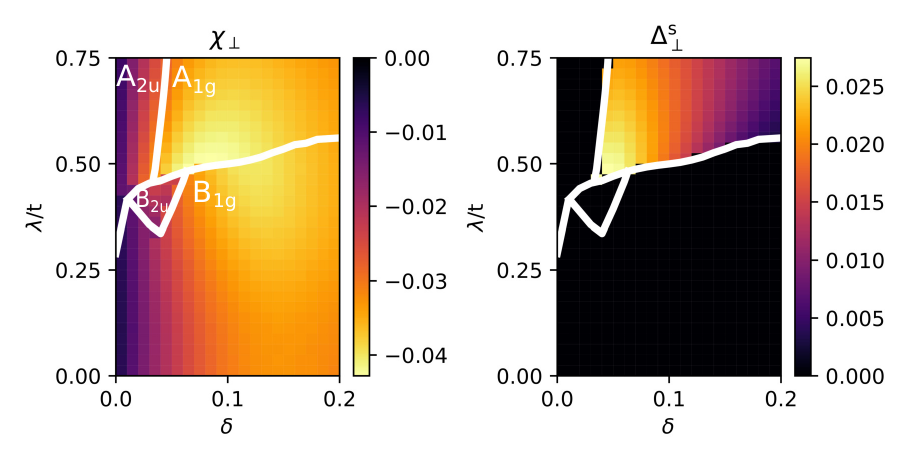}
	\caption{Interlayer MF parameter amplitudes for the bilayer system in $\delta-\lambda$ space for $t_\perp=0.2t$, irrep of mixed state realized is indicated. Note that the color ranges vary between plots for clarity.}
	\label{fig:OP_bi_ld}
\end{figure}

\Fref{fig:OP_bi_ld} shows the phase diagram in $\lambda-\delta$ space at fixed $t_\perp = 0.2t$. Only the variation of the interlayer MFs are shown, since the intralayer MFs varying negligibly from the monolayer results. The transition between the $d$-wave-like ($B_{1g}$ and $B_{2u}$) and the extended-$s$-wave-like ($A_{1g}$ and $A_{2u}$) states is very similar to the monolayer $B_1$ to $A_1$ transition. At low doping the odd-parity $A_{2u}$ and $B_{2u}$ states are more stable than their even-parity counterparts. Since singlet pairing is dominant in our model, this is equivalent a transition from zero to $\pi$ phase difference between the layers. This is remarkable, as the weak coupling of the layers should favour the zero phase difference, in analogy to a Josephson junction.


\subsection{Even-odd transitions}

\begin{figure}[t]
    \centering
    \begin{subfigure}{0.68\linewidth}
        \includegraphics[width=\textwidth]{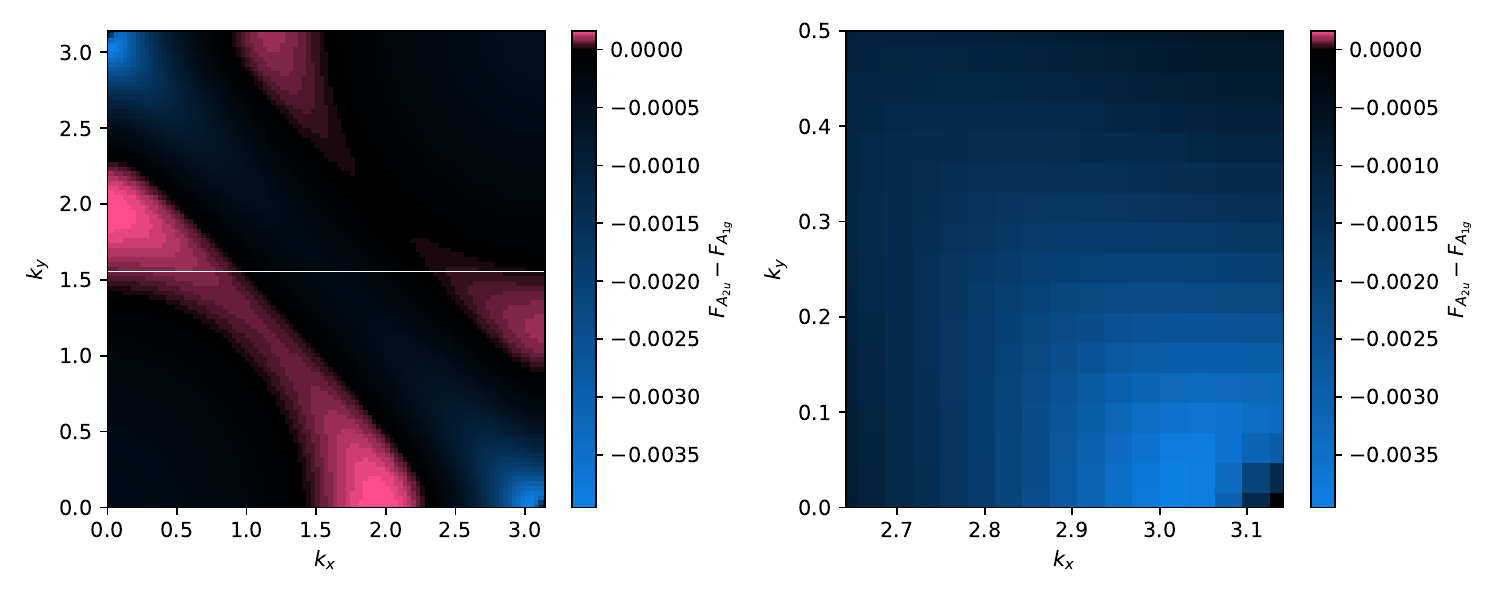}
        \caption{ }
    \end{subfigure}
    \begin{subfigure}{0.3\linewidth}
        \includegraphics[width=\textwidth]{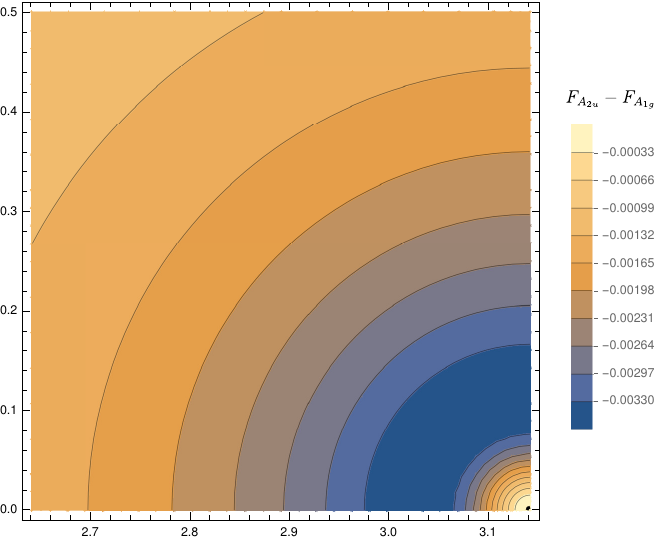}
        \caption{ }
    \end{subfigure}\\
    \begin{subfigure}{0.68\linewidth}
        \includegraphics[width=\textwidth]{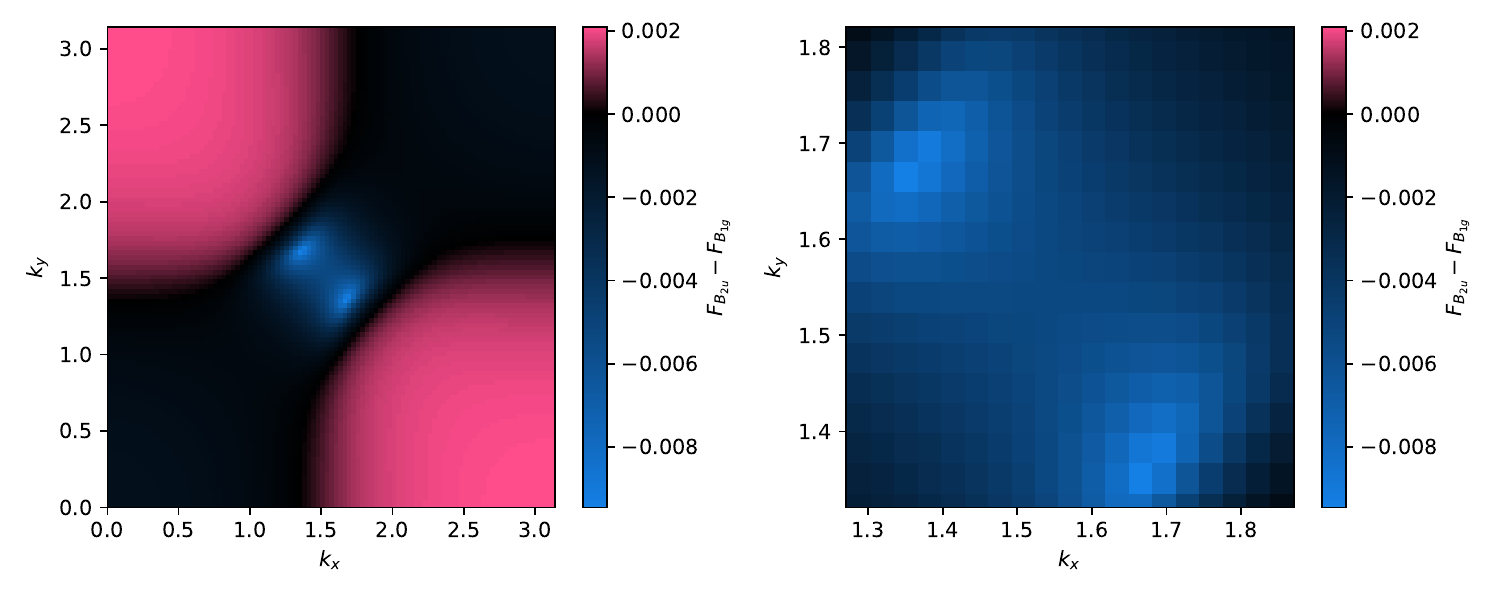}
        \caption{ }
    \end{subfigure}
    \begin{subfigure}{0.3\linewidth}
        \includegraphics[width=\textwidth]{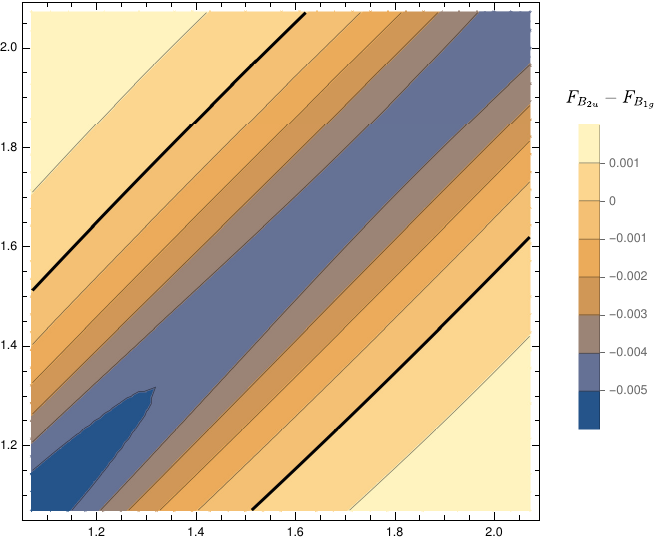}
        \caption{ }
    \end{subfigure}
    \caption{Free energy difference between the (a) $A_{2u}$ and $A_{1g}$ states in the bilayer with $t_\perp=0.2t$, $\lambda=0.6t$, $\delta=0.03$ with comparison to the first order expansion (b),  and (c) $B_{2u}$ and $B_{1g}$ states in the bilayer with $t_\perp=0.2t$, $\lambda=0.4t$, $\delta=0.03$ with comparison to the first order expansion (d). Negative values indicate stability of the odd state.}\label{fig:momdep}
\end{figure}

To investigate how the odd states are stabilized, the momentum space-resolved free energy difference was obtained to elucidate where the largest contributions to the stability of the odd states lie in momentum space. This is shown in \fref{fig:momdep}, where it is clear that the $A_{2u}$ is most stabilized about the $X$-point, whilst for the $B_{2u}$ this occurs about the point $k_x=k_y=\frac{\pi}{2}$.

\subsubsection{$A_{1g}$ and $A_{2u}$ irreps:}\label{sec:a1ga2u}

Firstly, for the $A_{1g}$/$A_{2u}$ irreps, we expand the Hamiltonian to first order in $(k_x,k_y)$ about the $X$-point, where the stabilization of the $A_{2u}$ is strongest. Note that NS terms which are prefactors to $\eta^\mu\otimes\sigma^\nu$ are denoted as the momentum-independent $\epsilon_{\mu\nu}$, and their SC counterparts prefixing $\eta^\mu\otimes\sigma^\nu(i\sigma^y)$ by $d_{\mu\nu}$, also momentum-independent. The shorthand $\epsilon_\lambda = \epsilon_{zx} = -\epsilon_{zy}$ will be used, and the symmetry of the different irreps will be utilized to simplify the algebra. For term containing the $A_{1g}$ triplet: $d_{zx}=-d_{zy}=d_t$ and 
\begin{align}
	\begin{split}
	H_{A_{1g}}^\text{BdG} =  &-\mu\eta^0\otimes\sigma^0 + \epsilon_{x0}\eta^x\otimes\sigma^0 + \epsilon_\lambda\eta^z\otimes\left(\sigma^xk_y - \sigma^yk_x\right) \\
	&+  d_{x0}\eta^x\otimes\sigma^0(i\sigma^y) + d_{t}\eta^z\otimes\left(\sigma^xk_y - \sigma^yk_x\right)(i\sigma^y),
	\end{split}
\end{align}
and likewise $d_{0x}=-d_{0y}=d_t$ for the $A_{2u}$:
\begin{align}
	\begin{split}
		H_{A_{2u}}^\text{BdG} =  &-\mu\eta^0\otimes\sigma^0 + \epsilon_{x0}\eta^x\otimes\sigma^0 + \epsilon_\lambda\eta^z\otimes\left(\sigma^xk_y - \sigma^yk_x\right) \\
		&+ d_{t}\eta^0\otimes\left(\sigma^xk_y - \sigma^yk_x\right)(i\sigma^y).
	\end{split}
\end{align}
The actual values for the $\epsilon_{\mu\nu}$ and $d_{\mu\nu}$ are very similar for the $A_{1g}$ and $A_{2u}$, with the obvious exception of the interlayer pairing. In the zero temperature limit, the electronic part of the energy of the ground state is given by the sum of the electron eigenenergies. This sum for the $A_{1g}$ is found to be
\begin{align}
    \begin{split}
        E_{A_{1g}} = &-2\sqrt{\mu ^2+\epsilon_{x0}^2  - 2\sqrt{\mu ^2 \left(\epsilon_\lambda^2 \tilde{q}^2+\epsilon_{x0}^2\right) + \left({d_t} \epsilon_{x0}-d_{x0} \epsilon_\lambda\right)^2\tilde{q}^2 }+\left({d_t}^2+\epsilon_\lambda^2\right) \tilde{q}^2+d_{x0}^2}\\
                    &-2\sqrt{\mu ^2+\epsilon_{x0}^2 +2 \sqrt{\mu ^2 \left(\epsilon_\lambda^2 \tilde{q}^2+\epsilon_{x0}^2\right)+ ({d_t} \epsilon_{x0}-d_{x0} \epsilon_\lambda)^2\tilde{q}^2}+\left({d_t}^2+\epsilon_\lambda^2\right) \tilde{q}^2+d_{x0}^2},
    \end{split}\label{eq:momdep_exp_a1g}
\end{align}
with $\tilde{q}^2=(k_x-\pi)^2+k_y^2$, and for the $A_{2u}$
\begin{align}
\begin{split}
    E_{A_{2u}} = &-\sqrt{\mu ^2+\epsilon_{x0}^2  -2 \sqrt{\mu ^2 \left(\epsilon_\lambda^2 \tilde{q}^2+\epsilon_{x0}^2\right)}+\left({d_t}^2+\epsilon_\lambda^2\right) \tilde{q}^2}\\
                    &-\sqrt{\mu ^2+\epsilon_{x0}^2+2 \sqrt{\mu ^2 \left(\epsilon_\lambda^2 \tilde{q}^2+\epsilon_{x0}^2\right)}+\left({d_t}^2+\epsilon_\lambda^2\right) \tilde{q}^2}.
\end{split}\label{eq:momdep_exp_a2u}
\end{align} 
This was found to be an acceptable level of approximation, as can be seen in \fref{fig:momdep}. We expand the energy difference $E_{A_{2u}}-E_{A_{1g}}$ to second order in $\tilde{q}$, which takes the form
\begin{align}
	E_{A_{2u}}-E_{A_{1g}} \approx P + Q\tilde{q}^2 + \ldots
\end{align}
with constants $P>0,\,Q<0$, and using the fact that the MF parameter values for both the $A_{1g}$ and the $A_{2u}$ are approximately equal. Constant $P$ arises from the interlayer pairing potential which is only present in the $A_{1g}$ state and, exactly at the $X$-point, lowers its energy compared to the $A_{2u}$ state. Since this $P$ is positive, the driving force for the stability of the $A_{2u}$ state must come from the momentum dependent terms. As the value of $\tilde{q}^2$ increases away from the $X$-point this term will bring the overall energy difference down and favour the $A_{2u}$, as is seen in \fref{fig:momdep}. Although the expression for $Q$ is very complicated, its negative value is due to a term proportional to $\left({d_t} \epsilon_{x0}-d_{x0} \epsilon_\lambda \right)^2$, so that if this term is finite then the $A_{2u}$ is stabilized in the vicinity of the $X$-point. Looking at the energies for the $A_{1g}$ \eref{eq:momdep_exp_a1g} and $A_{2u}$ \eref{eq:momdep_exp_a2u} it can be seen that this term is present in the expression for the $A_{1g}$ energy, but not that of the $A_{2u}$. 

We also find a link between the stability of the odd-parity state and the `superconducting fitness' \cite{fischer:2013,ramires:2016,ramires:2018}, which is a quantity providing a measure of the degree of destabilizing interband character in each pairing channel. In general, the fitness $F(\mathbf{k})$ is non-zero and given by
\begin{align}
    H_0(\mathbf{k})\Delta(\mathbf{k})-\Delta(\mathbf{k})H^\ast_0(-\mathbf{k}) = F(\mathbf{k})(i\sigma^y).
\end{align}
A smaller value of $\Tr\left\lbrace|F(\mathbf{k})|^2\right\rbrace$ indicates a more stable state, since this minimizes the degree of interband pairing. We calculate the fitnesses for both the even and odd gaps about the $X$-point using the approximate $H_0$:
\begin{align}
	\text{Tr}\left\lbrace \left(\Delta_{A_{1g}}H_0^T - H_0\Delta_{A_{1g}} \right)^2\right\rbrace &= -16(d_t\epsilon_{x0}-d_{x0}\epsilon_\lambda)^2\\
	\text{Tr}\left\lbrace \left(\Delta_{A_{2u}}H_0^T - H_0\Delta_{A_{2u}} \right)^2\right\rbrace &= 0.
\end{align}
This shows that about the $X$-point the $A_{2u}$ fitness is always smaller than the $A_{1g}$; in fact, with this vanishing trace, the $A_{2u}$ is said to be perfectly fit as it does not involve any interband pairing. Interestingly, the $A_{1g}$ result is proportional to the square of the extra factor appearing in the dispersion compared to the $A_{2u}$, mentioned before as the determining factor for $A_{2u}$ stability near the $X$-point. The size of this parameter shows a contradictory effect of the NS MF and the associated SC mean field, e.g. $d_t$ and $\epsilon_\lambda$. The full implications are hard to disentangle, but this is consistent with the relatively small region where the $A_{2u}$ phase is realized.

\subsubsection{$B_{1g}$ and $B_{2u}$ irreps:}

For the $B_{1g}$ and $B_{2u}$, the dispersions evaluated at the point $\left(\frac{\pi}{2},\frac{\pi}{2}\right)$ are:
\begin{align}
	E_{B_{1g},\pm} &= -2\left| \sqrt{\epsilon_{x0}^2+2\epsilon_\lambda^2} \pm \sqrt{2d_t^2+\mu^2} \right|\label{eq:b1g}\\
	E_{B_{2u},\pm} &= -2\sqrt{ 2d_t^2 + \epsilon_{x0}^2 + 2\epsilon_\lambda^2 + \mu^2 \pm 2\sqrt{4d_t^2\epsilon_\lambda^2 + (\epsilon_{x0}^2+2\epsilon_\lambda^2)\mu^2 }}\nonumber\\
				   &\approx -2\sqrt{ \left(\sqrt{\epsilon_{x0}^2 + 2\epsilon_\lambda^2 } \pm \mu\right)^2 + 2d_t^2\left(1\pm \frac{2\epsilon_\lambda^2}{\sqrt{\epsilon_{x0}^2+2\epsilon_\lambda^2}|\mu| } \right)}.
\end{align}
It can be seen that the $B_{1g}$ cannot open any effective gap at this point -- the gap simply renormalizes the chemical potential. This is indeed true for any point on the line $k_x=k_y$, due to the mirror antisymmetry along the diagonal for the $B_{1g}$ irrep. The $B_{2u}$, on the other hand has a gapped dispersion when the ILH term is nonzero, and this stabilizes the state relative to the $B_{1g}$. More explicitly: at the Fermi energy $\sqrt{\epsilon_{x0}^2 + 2\epsilon_\lambda^2 } = |\mu|$ holds, and if the ILH term $\epsilon_{x0}$ is taken to zero, $\sqrt{2\epsilon_\lambda^2 } = |\mu|$ and the gap in the $B_{2u}$ closes, becoming degenerate with the $B_{1g}$; the effective gap of the $B_{2u}$ is only opened by the ILH term. Furthermore, the point $\left(\frac{\pi}{2},\frac{\pi}{2}\right)$ lies on the nodal line for the $d$-wave singlet gap, so the effective gap is opened almost exclusively by the triplet channel. This argument is consistent with the numerically obtained dispersion, in which the gap near the point $\left(\frac{\pi}{2},\frac{\pi}{2}\right)$ closes only for the $B_{1g}$.

\subsection{Effect of an external magnetic field}

For a $\hat{z}$-aligned external magnetic field $H_Z$, the Zeeman Hamiltonian is
\begin{align}
	H_Z = \frac{-g\mu_B}{2} \sum_{kss'l} H_z \sigma_z c^\dagger_{ksl}c_{ks'l}.
\end{align}
Applying this field adds an $A_{2g}$ symmetry distortion in the bilayer, the same irrep as the $z$-rotation. This lowers the symmetry of the system, taking it from $D_{4h}$ to $C_{4h}$. This allows the $A_{1u}$, $A_{2g}$, $B_{1u}$, $B_{2g}$ to mix with the $A_{2u}$, $A_{1g}$, $B_{2u}$, $B_{1g}$ irreps respectively (\Tref{tab:biOPs}). The $\eta^y\sigma^z$ interlayer triplet in $A_u$ is sensitive only to the AFM interlayer coupling, with respect to which it is repulsive; as such it was found to have vanishing magnitude. Again, only the uniform extended $s$-wave dominant state ($A_g$) hosts an interlayer pairing. Of all new intralayer MFs introduced in \tref{tab:biOPs} only the additional triplet state in $A_g$ and $A_u$ was found to be nonzero. Nevertheless it is ignored in the following since the amplitude was negligible compared to the other MFs. 

\subsubsection{$A_{g}$ and $A_{u}$ irreps:}

\begin{figure}
    \centering
\begin{subfigure}{0.49\linewidth}
    \includegraphics[width=\linewidth]{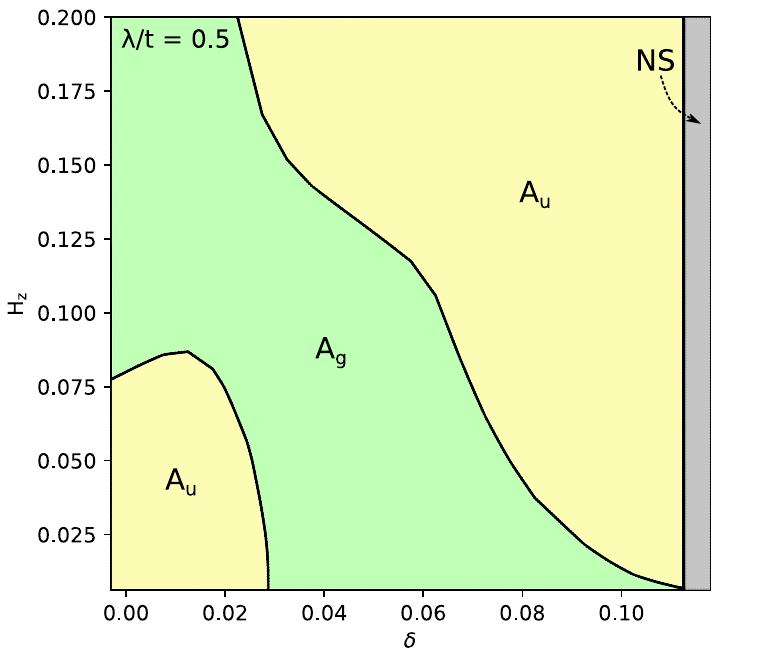}
    \subcaption{ }
    \label{fig:pd_HdA}
\end{subfigure}
\begin{subfigure}{0.49\linewidth}
    \includegraphics[width=\linewidth]{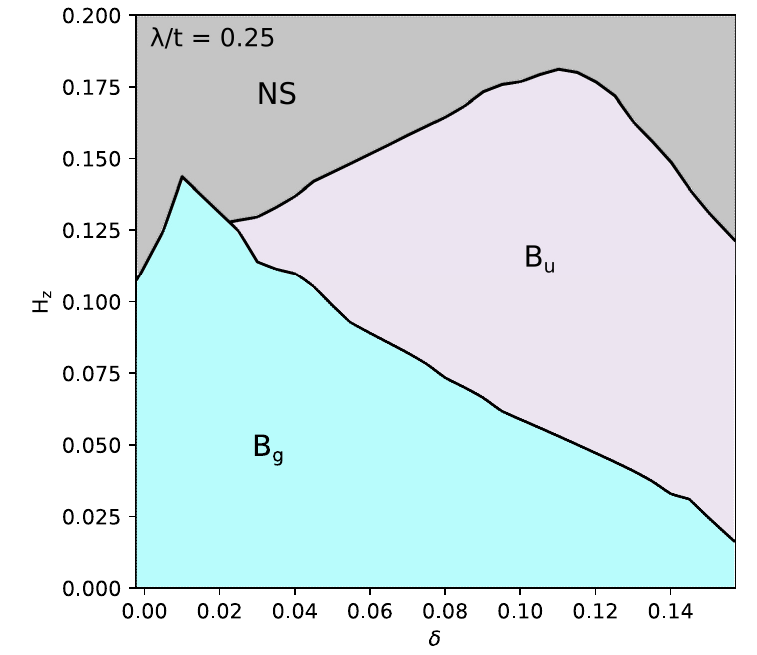}
    \subcaption{ }
    \label{fig:pd_HdB}
\end{subfigure}
\caption{Phase diagrams for the bilayer system in $\delta-H_z$ space at (a) $\lambda=0.5t$, $t_\perp=0.2t$ and (b) $\lambda=0.25t$, $t_\perp=0.2t$.}
\end{figure}

In \fref{fig:pd_HdA} at intermediate SOC ($\lambda=0.5t$), the $A_g$ state undergoes a field-induced first order transition out of the superconducting state at increasingly lower doping as the field strength increases, whereas the odd parity $A_u$ superconducting state is stable with respect to the applied magnetic field and experiences nearly no change in the MF magnitudes. There is a small region at low field/low doping where the odd parity state is more favoured, but with a much smaller free energy difference, although it is unclear whether this is simply due to proximity to the transition. The presence of the $A_u$ phase at low doping is not surprising given the presence of the $A_{2u}$ in the zero-field results, but it is unclear why it is suppressed by increasing magnetic field. Overall we see two possible parity switches within the superconducting state, seen in \fref{fig:pd_HdA}. Also present is the expected smooth second order transition into the NS with increased doping, and this is almost constant in $H_z$ owing to the stability of the $A_u$ state to the applied field.

\subsubsection{$B_{g}$ and $B_{u}$ irreps:}

For the $B_{g/u}$ irreps at $\lambda=0.25t$ the $B_g$ is clearly sensitive to the magnetic field and undergoes a first order transition into the NS at lower doping as the field strength is increased. The $B_u$ state persists to higher fields than the $B_g$ for given $\delta$, but as doping increases the $B_u$ state reaches its maximum critical field -- three times that of the $B_g$ -- at $\delta\approx0.11$, before being gradually suppressed at higher doping. The phase diagram comparing these two states at $\lambda=0.25t$ is seen in \fref{fig:pd_HdB}, and shows that we expect a parity switch within the superconducting state as with the $A_{g/u}$ states at higher SOC strength. It may be noted that a similar regime with parameters $\lambda=0.3t$, $t_\perp=0.1t$, and the weak coupling interaction strength $V_\text{int}=2.2t$, was investigated in in~\cite{yoshida:2012}. Whilst they didn't solve for the transition between different SC states, the qualitative shape of the SC region in $H-T$ space was also indicative of two-phase superconductivity.

\subsubsection{Pseudospin:}

This stability of the odd parity states with respect to the magnetic field is consistent with previous findings~\cite{khim:2021,maruyama:2012, sigrist:2014,yoshida:2012,yoshida:2014}, and can be explained by rewriting the states in a pseudospin basis of eigenvectors of the NS Hamiltonian, and observing that the $A_u$ and $B_u$ states become pseudospin triplets with $\mathbf{d_k}$ vector perpendicular to the pseudospin Zeeman field, making them immune to magnetic fields applied in the $\hat{z}$ direction. The $A_g$ and $B_g$ states on the other hand become pseudospin singlets and so will be unstable to the magnetic field. A derivation can be found in the Supplementary Material of~\cite{khim:2021}. 

\section{Conclusions}\label{sec:conc}

We have conducted an investigation into the properties of the superconducting state of a strongly-coupled Rashba mono- and bilayer over a wide internal parameter space and in an applied magnetic field. We found that the symmetry of the realized SC state depends strongly on the spin-orbit coupling strength and its interplay with both the doping rate and interlayer hopping, and the transition between extended-$s$- and $d$-wave dominant states a common feature of the mono- and bilayer cases, with the $s$-wave state prevailing with both strong spin-orbit coupling strength and strong interlayer hopping.

Interestingly the odd parity states in the bilayer were found to have regions where they are preferred at low doping and strong spin-orbit coupling. This was observed for both the $s$- and $d$-wave dominant mixed states, and results in parity switches within the SC state. We investigated the momentum dependent quasiparticle energies and identified regions of the BZ responsible for the stabilization of the odd states. We found that the $A_{2u}$ gap is fitter than the $A_{1g}$ gap close to the $X$-point, and this appears to be critical to the stability of the $A_{2u}$ state. This uncovered that a complex interplay of SOC, ILH, and both inter- and intralayer pairings may be the main stabilizing factor of the $A_{2u}$ state, and that the ILH protects the opening of an effective gap at the point $k_x=k_y=\pi/2$ for the $B_{2u}$ where the effective gap for the $B_{1g}$ closes.

The odd parity states were found to be resistant to an applied magnetic field relative to the even states, as has been previously observed for similar systems. This resulted in the parity of the superconducting state switching as the magnetic field is increased in strength. The $B_{g/u}$ system proved much less resilient to the applied field than the $A_{g/u}$, possibly due to reduction in the effective $g$-factor at high SOC strength.

Our findings have application in modelling quasi-2D materials or bulk materials with superconducting planes with strong electron correlations such as heavy fermion SCs and other strong coupled SCs which crystallize in the same structure.
\section*{References} 
\bibliography{bibliography.bib}

\end{document}